\RequirePackage{plautopatch} 
\documentclass[12pt]{article}
\usepackage[utf8]{inputenc}

\usepackage[dvipdfmx]{graphicx}
\usepackage{url}
\usepackage{listings,jvlisting}
\usepackage{amssymb}
\usepackage{tabularx}
\usepackage{amsmath}
\usepackage{mdframed}
\usepackage{graphicx}
\usepackage{setspace}
\usepackage{here}
\usepackage{authblk}
\usepackage{mathtools}
\usepackage{amsthm}
\usepackage{algorithmic}
\usepackage{algorithm}
\usepackage{arydshln}
\usepackage{bigints}
\usepackage{hyperref}
\def\al{{\alpha}}
\def\be{{\beta}}

\def\de{{\delta}}

\def\si{{\sigma}}

\def\beh{{\hat \be}}

\def\muh{{\widehat \mu}}

\def\x{{\text{\boldmath $x$}}}

\def\z{{\text{\boldmath $z$}}}

\def\xb{{\overline \x}}

\def\Nc{{\cal N}}

\begin{document}
\title{Dynamic Borrowing Method for Historical Information Using a Frequentist Approach for Hybrid Control Design}
\author[1,2]{Masahiro Kojima\footnote{Address: Biometrics Department, R\&D Division, Kyowa Kirin Co., Ltd.,
Otemachi Financial City Grand Cube, 1-9-2 Otemachi, Chiyoda-ku, Tokyo 100-004, Japan. Tel: +81-3-5205-7200 \quad
E-Mail: masahiro.kojima.tk@kyowakirin.com}}
\affil[1]{Kyowa Kirin Co., Ltd}
\affil[2]{The Institute of Statistical Mathematics}
\maketitle

\abstract{\noindent

Information borrowing from historical data is gaining attention in clinical trials of rare and pediatric diseases, where statistical power may be insufficient for confirmation of efficacy if the sample size is small. Although Bayesian information borrowing methods are well established, test-then-pool and equivalence-based test-then-pool methods have recently been proposed as frequentist methods to determine whether historical data should be used for statistical hypothesis testing. Depending on the results of the hypothesis testing, historical data may not be usable. This paper proposes a dynamic borrowing method for historical information based on the similarity between current and historical data. In our proposed method of dynamic information borrowing, as in Bayesian dynamic borrowing, the amount of borrowing ranges from 0\% to 100\%. We propose two methods using the density function of the t-distribution and a logistic function as a similarity measure. We evaluate the performance of the proposed methods through Monte Carlo simulations. We demonstrate the usefulness of borrowing information by reanalyzing actual clinical trial data.

}
\par\vspace{4mm}

\section{Introduction}\label{sec1}
The widespread use of real-world data has increased the use of historical data. The Food and Drug Administration issued draft guidelines entitled "Considerations for the Design and Conduct of Externally Controlled Trials for Drug and Biological Products" in 2023~\cite{FDA}. Clinical trials using external controls, such as historical data, are expected to accelerate drug development in cases where sample sizes may be small, such as for rare diseases and pediatrics. A hybrid control design has been proposed in which data are extrapolated from external targets to increase accuracy by reducing the sample size and lowering the allocation rate to the control group~\cite{pocock1976combination,hobbs2011hierarchical,zhu2020hybrid}.

Heterogeneity between current and historical data must be considered when borrowing information. Bayesian information borrowing methods consider heterogeneity. For example, Pocock~\cite{pocock1976combination} proposed a method for adjusting the amount of information borrowed according to the size and variability of the differences between control and historical control data by adding the distribution of the differences between these data to the posterior distribution of the current treatment and control group data. With the posterior distribution, the amount of information borrowed is reduced when the variance of the differences or the differences themselves are large. However, it is not easy to estimate the variability of the difference when historical data are scarce. A power prior method~\cite{ibrahim2015power,ibrahim2000power} was also proposed for assigning weights to current and historical control group data between 0 and 1. The power prior method is easy to apply because it only assigns weights to the likelihood function. However, because setting weights is difficult, a modified power prior~\cite{duan2006evaluating,neuenschwander2009note} was proposed to estimate weights from the data. When there are multiple historical data, a meta-analytic predictive (MAP) prior~\cite{neuenschwander2010summarizing} is used to account for heterogeneity among those data. The MAP prior changes the information scale according to heterogeneity, but does not account for heterogeneity between current and historical control data. Therefore, a robust MAP prior was proposed ~\cite{schmidli2014robust} for cases where the number of historical trials is low. 

For frequentist analysis, hypothesis testing is conducted on the basis of differences between current and historical control data. Depending on the results, historical control data are used or not used. The original test-then-pool (TTP) method~\cite{viele2014use} and equivalence-based TTP method~\cite{li2020revisit} were proposed to this end. The problem with the two methods is that the historical data may not be usable depending on the results of the test because the decision is made by hypothesis testing.

We propose a dynamic borrowing method based on the degree of similarity between current and historical data. Our proposed dynamic information borrowing is a continuous quantity between $[0,1]$, where 1 indicates the borrowing of historical control data. We propose two different methods for calculating similarity. The first uses a t-distribution and the second uses a logistic function. Through Monte Carlo simulations, we evaluate the performance of our proposed methods. Finally, we apply our proposed methods to real study data and provide the reanalysis results.

This paper is organized as follows. Section 2 introduces the t-distribution- and logistic function-based methods. Section 3 describes the setting and results of the computer simulations. Section 4 presents the results for an actual clinical trial. Finally, we conclude the paper with a discussion in Section 5.

\section{Methods}\label{sec2}
We assume that the current data $X_t$ of the treatment group are normally distributed $\Nc(\mu_t,\si^2_t)$, and that $X_c$ of the control group is distributed as $\Nc(\mu_c,\si^2_c)$. The historical data $X_h$ of the control group are distributed as $\Nc(\mu_h,\si^2_h)$. We assume that there is one historical trial. The sample size is $n_t$ for the treatment group, $n_c$ for the current control group, and $n_h$ for the historical control group. The mean and standard deviation of the treatment group are $\xb_t$ and $s_t$, the mean and standard deviation of the control group for the current trial are $\xb_c$ and $s_c$, and the mean and standard deviation for the historical data are $\xb_h$ and $s_h$.

The primary hypothesis is 
\begin{align}
\label{eq:null}
H_0: \mu_t = \mu_{pc} \mbox{ vs. } H_1: \mu_t \neq \mu_{pc},
\end{align}
where $\mu_{pc}$ is the pooled mean of the current and historical control groups. 

\subsection{Review of previous borrowing methods applied to historical control data using the frequentist approach}
The borrowing of historical information in the frequentist approach involves the pooled mean and standard deviation for the current and historical control groups. We assume that the information-borrowing level is $a\in \{0,1\}$. The primary test statistic is expressed as
\begin{align}
T(a)=\cfrac{\xb_t-\cfrac{n_c\xb_c+an_h\xb_h}{n_c+an_h}}{\sqrt{\cfrac{s_t^2}{n_t}+\cfrac{n_cs_c^2+a^2n_hs_h^2}{(n_c+an_h)^2}}}.
\end{align}
When the test statistic $|T(a)|$ exceeds an upper $1-\frac{\al}{2}$ limit of the normal distribution, where $\al$ is a prespecified significance level, the null hypothesis is rejected. Here, we introduce a way to determine the value of $a$ for the original and equivalence-based TTP methods.

\subsubsection{Original test-then-pool method}
The original TTP method is a simple method that compares current control and historical control data. The method uses all of the historical data if there is no statistically significant difference between the current and historical control data, but only the current data if a significant difference is detected. The null hypothesis for the decision to use historical control data is as follows:
\begin{align}
H_{0h_1}: \mu_c=\mu_h\mbox{ vs.}H_{1h_1}: \mu_c\neq\mu_h.
\end{align}
The test statistic is 
\begin{align}
T_1=\frac{\xb_c-\xb_h}{\sqrt{\frac{s^2_c}{n_c}+\frac{s^2_h}{n_h}}}.
\end{align}

If $|T_1|$ exceeds $z_{1-\al_{h1}/2}$ ($1-\al_{h1}/2$ percentile points of the standard normal distribution), where $\al_{h1}$ is a prespecified significance level (e.g., 0.05 or 0.10), we apply $T(0)$ to verify the primary hypothesis ($\mu_{pc}$ becomes $\mu_c$ because the historical control data are not pooled.). If $|T_1|<z_{1-\al_{h1}/2}$ holds, we apply $T(1)$ to verify the primary hypothesis. In other words, the primary hypothesis is confirmed for pooled and non-pooled cases. Hence, the overall significance level needs to be adjusted. The adjusted significance level $\al^\ast_{TTP}$ under $H_0$ satisfies
\begin{align}
\label{eq:TTP_alpha}
    \al&=P\left(|T(1)|>z_{1-\al^\ast_{TTP}/2}-\frac{\frac{n_h(\mu_c-\mu_h)}{n_c+n_h}}{\sqrt{\frac{\si^2_t}{n_t}+\frac{n_c\si^2_c+n_h\si_h^2}{(n_c+n_h)^2}}},|T_1|\leq z_{1-\al_{h_1}/2}|H_0\right)\nonumber\\
    &\hspace{0.5cm}+P\left(|T(0)|>z_{1-\al^\ast_{TTP}/2},|T_1|>z_{1-\al_{h_1}/2}|H_0\right)\nonumber\\
    &=P\left(|T(1)|>z_{1-\al^\ast_{TTP}/2}-\frac{\frac{n_h(\mu_c-\mu_h)}{n_c+n_h}}{\sqrt{\frac{\si^2_t}{n_t}+\frac{n_c\si^2_c+n_h\si_h^2}{(n_c+n_h)^2}}},|T_1|\leq z_{1-\al_{h_1}/2}|H_0\right)\nonumber\\
    &\hspace{0.5cm}+P\left(|T(0)|>z_{1-\al^\ast_{TTP}/2}|H_0\right)-P\left(|T(0)|>z_{1-\al^\ast_{TTP}/2},|T_1|<z_{1-\al_{h_1}/2}|H_0\right)\nonumber\\
    &=2\int^{\infty}_{z_{1-\al^\ast_{TTP}/2}-\frac{\frac{n_h(\mu_c-\mu_h)}{n_c+n_h}}{\sqrt{\frac{\si^2_t}{n_t}+\frac{n_c\si^2_c+n_h\si_h^2}{(n_c+n_h)^2}}}}\int^{z_{1-\al_{h1}/2}\sqrt{\si^2_h/n_h+\si^2_c/n_c}}_{-z_{1-\al_{h1}/2}\sqrt{\si^2_h/n_h+\si^2_c/n_c}}f_{Y_1,Y_2}(y_1,y_2)dy_1dy_2\nonumber\\
    &\hspace{0.5cm}+\al^\ast_{TTP}-2\int^{\infty}_{z_{1-\al^\ast_{TTP}/2}}\int^{z_{1-\al_{h1}/2}\sqrt{\si^2_h/n_h+\si^2_c/n_c}}_{-z_{1-\al_{h1}/2}\sqrt{\si^2_h/n_h+\si^2_c/n_c}}f_{Y_3,Y_2}(y_3,y_2)dy_3dy_2
\end{align}
where $\alpha$ is a nominal significance level (eg. 0.05) and the density function is $f_{Y_1,Y_2}(y_1,y_2)$ according to the bivariate normal distribution of $(Y_1,Y_2)$
\begin{align}
\begin{pmatrix}
 Y_1 \\
 Y_2
\end{pmatrix}
\sim\Nc
\begin{pmatrix}
\begin{pmatrix}
0\\
\mu_c-\mu_h
\end{pmatrix}
,
\begin{pmatrix}
1 & \frac{\frac{\si_h^2-\si_c^2}{n_c+n_h}}{\sqrt{\frac{\si_t^2}{n_t}+\frac{n_c\si_c^2+n_h\si_h^2}{(n_c+n_h)^2}}}\\
\frac{\frac{\si_h^2-\si_c^2}{n_c+n_h}}{\sqrt{\frac{\si_t^2}{n_t}+\frac{n_c\si_c^2+n_h\si_h^2}{(n_c+n_h)^2}}} & \si_h^2/n_h+\si_c^2/n_c
\end{pmatrix}
\end{pmatrix},
\end{align}
and the density function is $f_{Y_3,Y_2}(y_3,y_2)$ according to the bivariate normal distribution of $(Y_3,Y_2)$
\begin{align}
\begin{pmatrix}
 Y_3 \\
 Y_2
\end{pmatrix}
\sim\Nc
\begin{pmatrix}
\begin{pmatrix}
0\\
\mu_c-\mu_h
\end{pmatrix}
,
\begin{pmatrix}
1 & -\frac{\frac{\si_c^2}{n_c}}{\sqrt{\frac{\si_t^2}{n_t}+\frac{\si_c^2}{n_c}}}\\
-\frac{\frac{\si_c^2}{n_c}}{\sqrt{\frac{\si_t^2}{n_t}+\frac{\si_c^2}{n_c}}} & \si_h^2/n_h+\si_c^2/n_c
\end{pmatrix}
\end{pmatrix}
.
\end{align}

For either $T(0)$ or $T(1)$, the primary hypothesis testing of (\ref{eq:null}) is conducted using an upper $\al^\ast_{TTP}/2$ threshold of the standard normal distribution.

\subsection{Equivalence-based test-then-pool method}
The equivalence-based TTP method differs from the original TTP method in that the use of historical data is determined by equivalence-based testing.
\begin{align}
H_{0h_2}: \mu_c-\mu_h\leq\de\mbox{ or }\mu_c-\mu_h\geq-\de\mbox{ vs. }H_{1h_2}: -\de<\mu_c-\mu_h<\de,
\end{align}
where $\de$ is an equivalence margin. When the distribution of the current control group is $\Nc(\mu_c,1)$ and the distribution of the historical control group is $\Nc(\mu_c\pm \de,1)$, $\de$ is 0.25 or 0.30 and there is overlap between the distribution of the current and historical control group of about 90\% \cite{li2020revisit}. If the standard deviation is not 1, the margin $\de$ is $0.25\si$ or $0.30\si$ when $\si_c$=$\si_h$=$\si$. For an explanation of how to proceed if the variances are not equal, see section 3.1 of Li et al\cite{li2020revisit}.

$T(1)$ is used if the following situation is satisfied.
\begin{align}
    -\frac{\de}{\sqrt{\frac{s^2_c}{n_c}+\frac{s^2_h}{n_h}}}+Z_{1-\al_2}<T_1<\frac{\de}{\sqrt{\frac{s^2_c}{n_c}-\frac{s^2_h}{n_h}}}-Z_{1-\al_2}
\end{align}
If not satisfied, $T(0)$ is used.

 For primary hypothesis testing, historical data may or may not be pooled. Hence, the significance level needs to be adjusted. The adjusted significance level $\al^\ast_{EQ}$ satisfies
\begin{align}
\label{eq:EQ_alpha}
    \al&=P\left(|T(1)|>z_{1-\al^\ast_{EQ}/2}-\frac{\frac{n_h(\mu_c-\mu_h)}{n_c+n_h}}{\sqrt{\frac{\si^2_t}{n_t}+\frac{n_c\si^2_c+n_h\si_h^2}{(n_c+n_h)^2}}},|T_1|\leq \frac{\de}{\sqrt{\si^2_h/n_h+\si^2_c/n_c}}-z_{1-\al_{h_2}}|H_0\right)\nonumber\\
    &\hspace{0.5cm}+P\left(|T(0)|>z_{1-\al^\ast_{EQ}/2},|T_1|>\frac{\de}{\sqrt{\si^2_h/n_h+\si^2_c/n_c}}-z_{1-\al_{h_2}}|H_0\right)\nonumber\\
    &=P\left(|T(1)|>z_{1-\al^\ast_{EQ}/2}-\frac{\frac{n_h(\mu_c-\mu_h)}{n_c+n_h}}{\sqrt{\frac{\si^2_t}{n_t}+\frac{n_c\si^2_c+n_h\si_h^2}{(n_c+n_h)^2}}},|T_1|\leq \frac{\de}{\sqrt{\si^2_h/n_h+\si^2_c/n_c}}-z_{1-\al_{h_2}}|H_0\right)\nonumber\\
    &\hspace{0.5cm}+P\left(|T(0)|>z_{1-\al^\ast_{EQ}/2}|H_0\right)-P\left(|T(0)|>z_{1-\frac{\al^\ast_{TTP}}{2}},|T_1|<\frac{\de}{\sqrt{\si^2_h/n_h+\si^2_c/n_c}}-z_{1-\al_{h_2}}|H_0\right)\nonumber\\
    &=\frac{\al^\ast_{EQ}}{2}+\int^{\infty}_{\z_{1-\al^\ast_{EQ}}-\frac{\frac{n_h(\mu_c-\mu_h)}{n_c+n_h}}{\sqrt{\frac{\si^2_t}{n_t}+\frac{n_c\si^2_c+n_h\si_h^2}{(n_c+n_h)^2}}}}\int^{\de-z_{1-\al_{h2}}\sqrt{\si^2_h/n_h+\si^2_c/n_c}}_{-\de+z_{1-\al_{h2}}\sqrt{\si^2_h/n_h+\si^2_c/n_c}}f_{Y_1,Y_2}(y_1,y_2)dy_1dy_2\nonumber\\
    &\hspace{0.5cm}-\int^{\infty}_{z_{1-\al^\ast_{EQ}}}\int^{\de-z_{1-\al_{h2}}\sqrt{\si^2_h/n_h+\si^2_c/n_c}}_{-\de+z_{1-\al_{h2}}\sqrt{\si^2_h/n_h+\si^2_c/n_c}}f_{Y_3,Y_2}(y_3,y_2)dy_3dy_2
\end{align}

For both $T(0)$ and $T(1)$, primary hypothesis testing of (\ref{eq:null}) is conducted using an upper $\al^\ast_{EQ}/2$ threshold of the standard normal distribution.

When the null hypothesis $H_{0h_1}$ is rejected for the original TTP method, we cannot fully use the historical control design; this is also the case when the null $H_{0h_2}$ is not rejected for the equivalence-based TTP method. With these methods, the amount of information borrowed from the historical data varies continuously between 0 and 1, depending on the similarity between the current and historical control data, as shown in the next section.

\subsection{Dynamic historical data borrowing method}
We propose a dynamic borrowing method for historical information. In fact, we propose two methods in which the amount of information borrowing is dynamically substituted depending on the data. Because $T_1$ follows a t-distribution, we define the dynamic historical data borrowing level on the basis of a density function according to the following t-distribution: 
\begin{align}
    A_t(\xb_c,s_c,\xb_h,s_h)=\frac{f_t\left(|T_1|\right)}{f_t(0)}.
\end{align}
where $t$ is a random variable with a t-distribution $t_{n_c+n_h-2}$. $A_t(\xb_c,s_c,\xb_h,s_h)$ takes values of $[0,1]$. We define the dynamic borrowing test statistics $T(A_t(\xb_c,s_c,\xb_h,s_h))$. Because the asymptotic distribution of $T(A_t(\xb_c,s_c,\xb_h,s_h))$ is difficult to derive, hypothesis testing is carried out using parametric bootstrap samples. For the bootstrap-based hypothesis testing, first, $B$ sets of bootstrap samples are generated by $\Nc(\muh,s^2_t)$, $\Nc(\muh,s^2_c)$, and $\Nc(\muh,s^2_h)$, respectively, where $\muh$ has no problem with any value (e.g., 0). Let the bootstrap test statistic of the $b$-th set of bootstrap samples be $T^\ast_b(A_t(\xb^\ast_{bc},s^\ast_{bc},\xb^\ast_{bh},s^\ast_{bh}))$ $(b=1,2,\ldots,B)$. The critical value $z^\ast_{1-\al/2\%}$ for the hypothesis testing is the upper $\al/2\%$-th percentile of the $B$ bootstrap test statistics. The p-value is the proportion of the $B$ bootstrap test statistics exceeding the test statistic $A_t(\xb_c,s_c,\xb_h,s_h)$. In other words, the one sided p-value is$=\frac{1}{B}\sum^B_{b=1}I(A_t(\xb_c,s_c,\xb_h,s_h)<T^\ast_b(A_t(\xb^\ast_{bc},s^\ast_{bc},\xb^\ast_{bh},s^\ast_{bh})))$, where $I(\cdot)$ is an indicator function. For the dynamic borrowing method based on the density function of the t-distribution, there is insufficient borrowing of historical control data. For example, the sample size of both the current and historical control groups is 50, and the amount of borrowed information at the upper $5\%$ percentile point of the t-distribution is 0.25, which may be too small. In this case, we consider another method in which information borrowing decreases slowly until a certain point and decreases rapidly from a point at which similarity is considered small.

Another dynamic borrowing method uses the following logistic function.
\begin{align}
    A_l(\xb_c,s_c,\xb_h,s_h)=\frac{1}{1+\exp(\be_0+\be_1|T_1|)}
\end{align}
This method requires an estimation of $\be_0$ and $\be_1$. Because the logistic function does not fit to some data, the parameters are arbitrarily determined; for example, the information borrowed is 0.5 when $T_1$ is 1.64 (upper 5\% point of the standard normal distribution) and 0.2 when $T_1$ is 1.96 (upper 2.5\% point of the standard normal distribution), then $\beh_0=-7.379$ and $\beh_1=4.472$. As another example, the information borrowed is 0.5 when $T_1$ is 1.96 (upper 2.5\% point of the standard normal distribution) and 0.2 when $T_1$ is 2.33 (upper 1\% point of the standard normal distribution), then $\beh_0=-7.374$ and $\beh_1=3.747$. $A_l(\xb_c,s_c,\xb_h,s_h)$ does not hold 1 regardless of $\beh_0=-\infty$. However, because the examples above show that the amount of borrowed information is 0.9994 at $T_1=0$, we consider that there is no problem. This method was also tested using the bootstrap method because it is difficult to derive the asymptotic distribution of $T_l(A(\xb_c,s_c,\xb_h,s_h))$.

From here, we call the dynamic borrowing information based on the density function of the t-distribution DB-T, the dynamic borrowing information based on the logistic function with $\beh_0=-7.379$ and $\beh_1=4.472$ DB-L1, the dynamic borrowing information based on the logistic function with $\beh_0=-7.374$ and $\beh_1=3.747$ DB-L2, the original TTP method TTP, and the equivalence-based TTP method EQ.

\begin{figure}[H]
  \begin{center}
  \includegraphics[width=15cm]{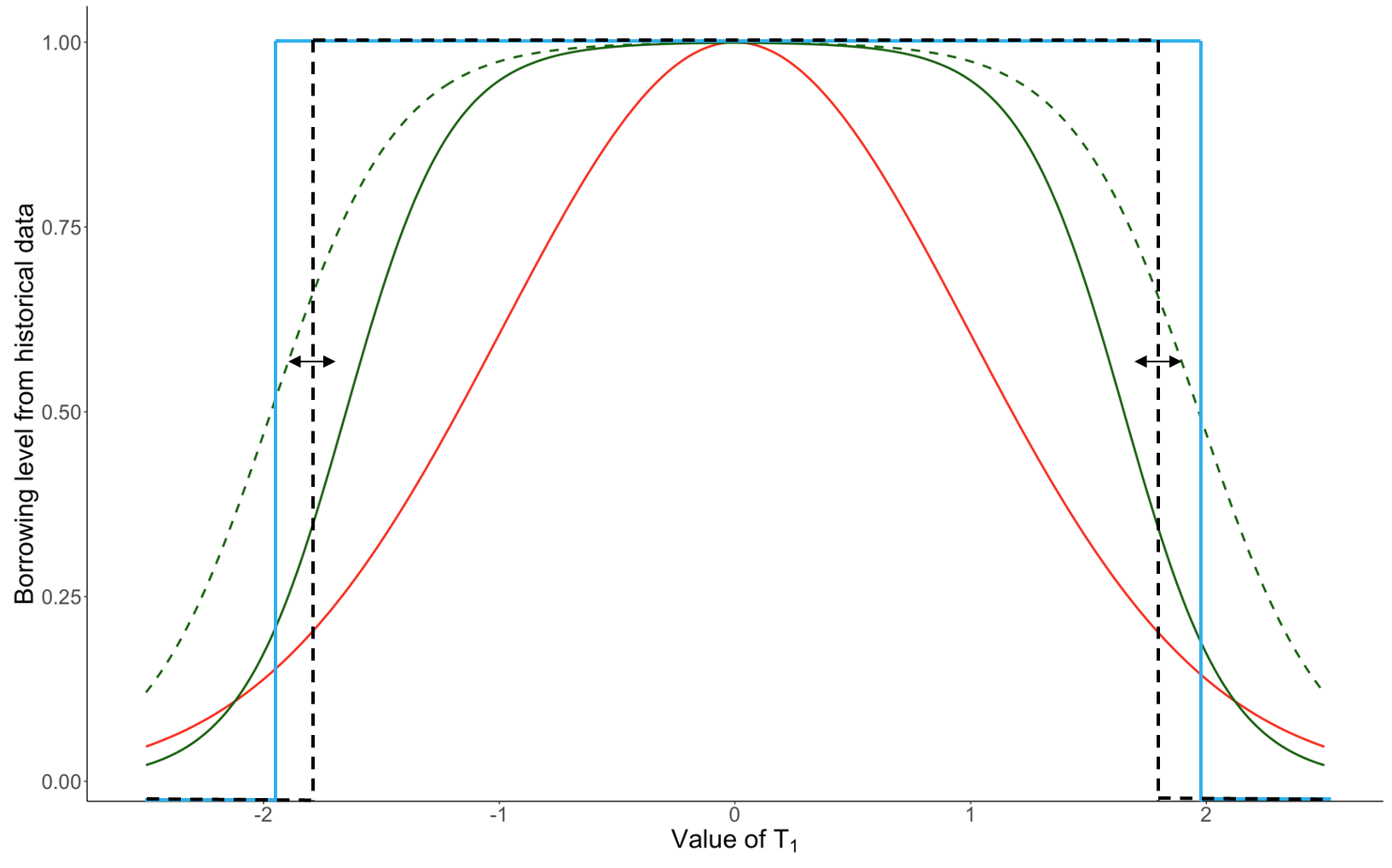}
  \caption{borrowing level}
  \label{f:weight}
      \footnotesize{The red line is our proposed method based on $t_{98}$, the green line is our proposed method based on the logistic function with $\beh_0=-7.379$ and $\beh_1=4.472$, the green dashed line is our proposed method based on the logistic function with $\beh_0=-7.374$ and $\beh_1=3.747$, the light blue line is the TTP method with $\al_{h1}=0.05$, and the black dashed line is the equivalence-based TTP method (the results of the equivalence-based TTP method vary with sample size, standard deviation, and $\de$)}
  \end{center}
\end{figure}

We show the amount of information borrowing in Figure \ref{f:weight} within $T_1\in[-2.5,2.5]$ under $n_c=50$ and $n_h=50$. On the basis of the t-distribution, the total amount of dynamic borrowing is smaller than in the other methods, at 1.96, in which the amount of borrowing in TTP is 0 is 0.14. When applying the logistic function, the amount of borrowing is larger than when using the method based on the t-distribution. Especially, the amount is almost $1$ when $T_1$ is around $0$. In addition, in the area where the amount of borrowing in TTP becomes $0$, some information is borrowed. For the EQ, the range of $T_1$ where the amount of borrowing is 1 is similar to that with TTP. However, the range of EQ can change depending the sample size, standard deviation, and $\de$. For example, under the situation where the standard deviation is $5$ and $\de=0.3$, the amount of borrowing is 1 within $T_1 \in [-1.71,1.71]$ when $n_c=n_h=50$, and is 1 within $T_1 \in [-3.10,3.10]$ when $n_c=n_h=100$.

\section{Simulation study}
We evaluate the performance of our proposed methods by comparing the original TTP and EQ methods using Monte Carlo simulations. Specifically, the simulation evaluated DB-T, dynamic borrowing based on the logistic function with $\be_0=-7.379$ and $\be_1=4.472$ (DB-L1), and dynamic borrowing based on the logistic function with $\be_0=-7.374$ and $\be_1=3.747$ (DB-L2).

\subsection{Simulation configuration}
The true values of the main parameters are listed in Table \ref{t:para}. Scenarios 1-4 aim to confirm that the one-sided type 1 error maintains a nominal significance level of 2.5\%. The purpose of Scenarios 5-24 is to compare statistical power among the methods. The number of simulations in each scenario is 100,000. The number of bootstrap sets is 10,000. The type 1 error in the TTP method is adjusted by the equation (\ref{eq:TTP_alpha}) under the sample sizes and standard deviations shown in Table \ref{t:para}. The equivalence margin $\de$ is $1.5$, satisfying the overlap area of about 90\%. The type 1 error in the EQ method is adjusted by the equation (\ref{eq:EQ_alpha}) under the sample sizes and standard deviations shown in Table \ref{t:para} and $\de=1.5$. For $\al_{h1}$ and $\al_{h2}$, we use 0.05 and 0.10, respectively. We call TTP1 under $\al_{h1}=0.05$, TTP2 under $\al_{h1}=0.10$, EQ1 under $\al_{h2}=0.05$, and EQ2 under $\al_{h2}=0.10$.

\begin{table}[H]
  \begin{center}
\caption{Parameter setting\label{t:para}}
\begin{tabular}{|c|c|c|c|c|c|c|c|}\hline
Scenario& $n_t$ & $n_c$&$n_h$ &$\mu_t$ & $\mu_c$ & $\mu_h$ & $\si_t^2=\si_c^2=\si_h^2$\\ \hline
1& 50 & 25 & 25 & 0 & 0 & 0 & 5\\
2& 50 & 50 & 50 & 0 & 0 & 0 & 5\\
3& 100 & 50 & 50 & 0 & 0 & 0 & 5\\
4& 100 & 100 & 100 & 0 & 0 & 0 & 5\\\hline
5& 50 & 25 & 25 & 2 & 0 & 0 & 5\\
6& 50 & 50 & 50 & 2 & 0 & 0 & 5\\
7& 100 & 50 & 50 & 2 & 0 & 0 & 5\\
8& 100 & 100 & 100 & 2 & 0 & 0 & 5\\\hline
9& 50 & 25 & 25 & 2 & 0 & 1 & 5\\
10& 50 & 50 & 50 & 2 & 0 & 1 & 5\\
11& 100 & 50 & 50 & 2 & 0 & 1 & 5\\
12& 100 & 100 & 100 & 2 & 0 & 1 & 5\\\hline
13& 50 & 25 & 25 & 2 & 0 & -1 & 5\\
14& 50 & 50 & 50 & 2 & 0 & -1 & 5\\
15& 100 & 50 & 50 & 2 & 0 & -1 & 5\\
16& 100 & 100 & 100 & 2 & 0 & -1 & 5\\\hline
17& 50 & 25 & 25 & 2 & 1 & 0 & 5\\
18& 50 & 50 & 50 & 2 & 1 & 0 & 5\\
19& 100 & 50 & 50 & 2 & 1 & 0 & 5\\
20& 100 & 100 & 100 & 2 & 1 & 0 & 5\\\hline
21& 50 & 25 & 25 & 2 & -1 & 0 & 5\\
22& 50 & 50 & 50 & 2 & -1 & 0 & 5\\
23& 100 & 50 & 50 & 2 & -1 & 0 & 5\\
24& 100 & 100 & 100 & 2 & -1 & 0 & 5\\\hline
\end{tabular}
  \end{center}
\end{table}

\subsection{Simulation results}
\begin{figure}[H]
  \begin{center}
  \includegraphics[width=15cm]{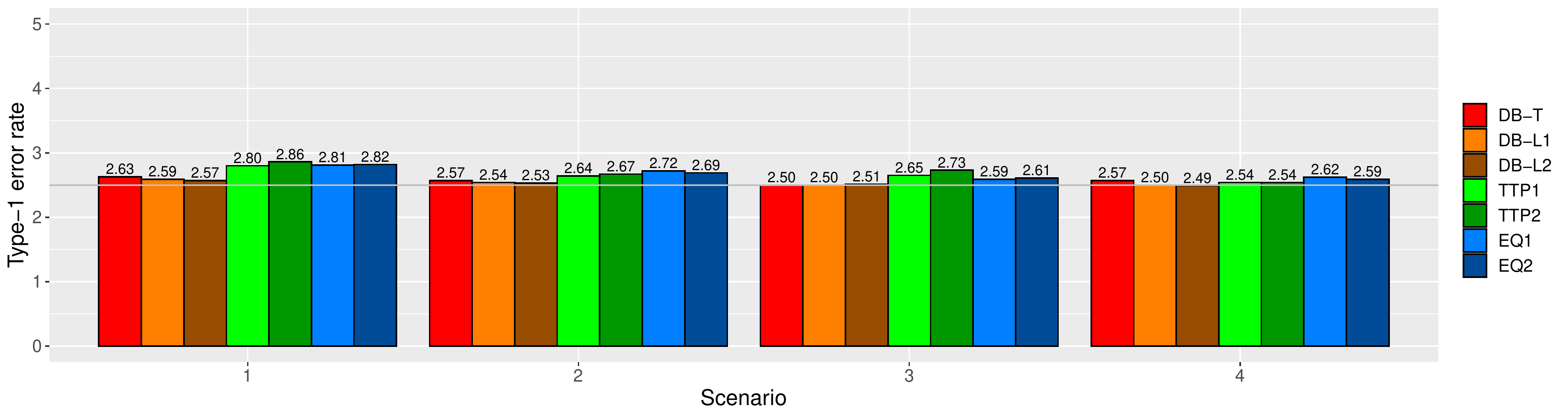}
  \caption{type 1 error}
  \label{f:type1error}
     \footnotesize{DB-T is the dynamic borrowing method based on the density function of the t-distribution, DB-L1 is the dynamic borrowing based on the logistic function with $\be_0=-7.379$ and $\be_1=4.472$, DB-L2 is the dynamic borrowing based on the logistic function with $\be_0=-7.374$ and $\be_1=3.747$, TTP1 is the original TTP method under $\al_{h1}=0.05$, TTP2 is the original TTP method under $\al_{h1}=0.10$, EQ1 is the equivalence-based TTP method under $\al_{h2}=0.05$, and EQ2 is the equivalence-based TTP method under $\al_{h2}=0.10$, Gray line is 2.5\%}
  \end{center}
\end{figure}

The type 1 errors for Scenarios 1-4 are shown in Figure \ref{f:type1error}. The type 1 errors of DB-T range from 2.63 to 2.50. The type 1 errors of DB-L1 and DB-L2 range from of 2.59 to 2.49. The type 1 errors of TTP1 and TTP2 range from 2.86 to 2.54. The type 1 errors of EQ1 and EQ2 range from 2.82 to 2.59. 

\begin{figure}[H]
  \begin{center}
  \includegraphics[width=15cm]{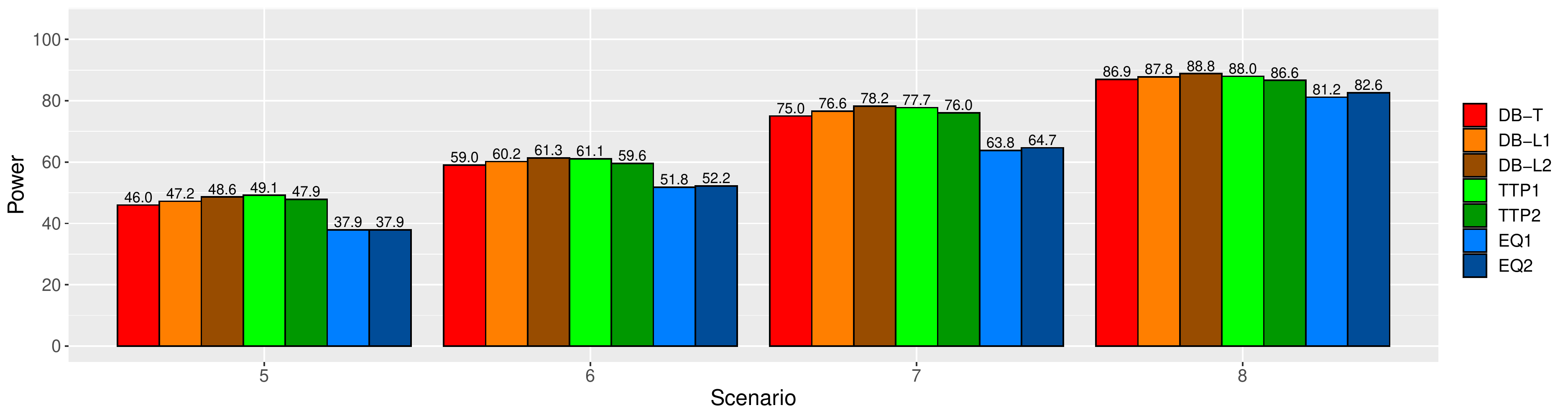}
  \caption{Power in Scenarios 5-8}
  \label{f:power1}
     \footnotesize{DB-T is the dynamic borrowing method based on the density function of the t-distribution, DB-L1 is the dynamic borrowing based on the logistic function with $\be_0=-7.379$ and $\be_1=4.472$, DB-L2 is the dynamic borrowing based on the logistic function with $\be_0=-7.374$ and $\be_1=3.747$, TTP1 is the original TTP method under $\al_{h1}=0.05$, TTP2 is the original TTP method under $\al_{h1}=0.10$, EQ1 is the equivalence-based TTP method under $\al_{h2}=0.05$, and EQ2 is the equivalence-based TTP method under $\al_{h2}=0.10$}
  \end{center}
\end{figure}

\begin{figure}[H]
  \begin{center}
  \includegraphics[width=15cm]{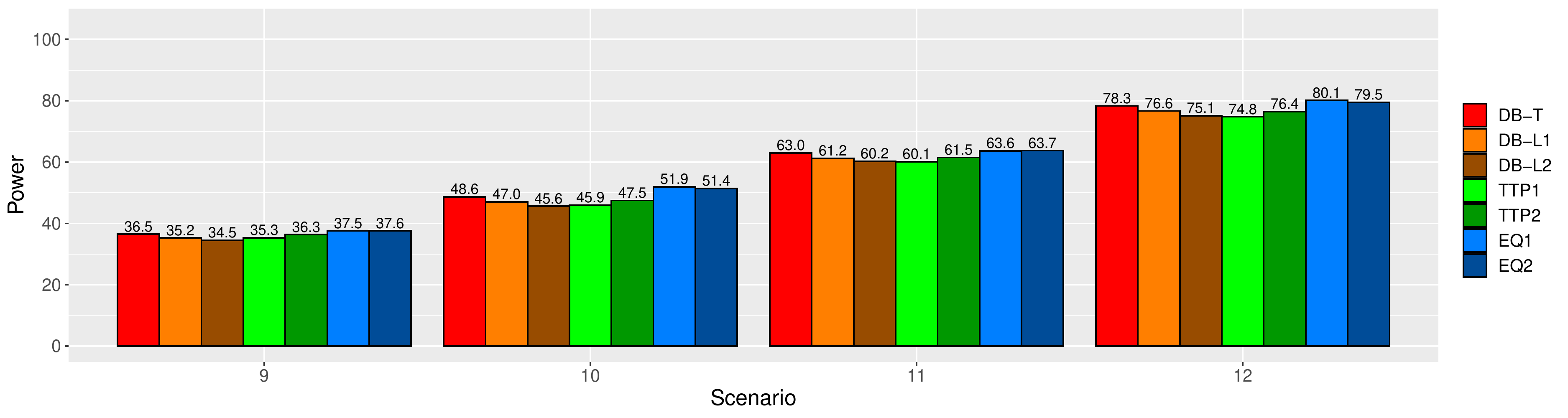}
  \caption{Power in Scenarios 9-12}
  \label{f:power2}
     \footnotesize{DB-T is the dynamic borrowing method based on the density function of the t-distribution, DB-L1 is the dynamic borrowing based on the logistic function with $\be_0=-7.379$ and $\be_1=4.472$, DB-L2 is the dynamic borrowing based on the logistic function with $\be_0=-7.374$ and $\be_1=3.747$, TTP1 is the original TTP method under $\al_{h1}=0.05$, TTP2 is the original TTP method under $\al_{h1}=0.10$, EQ1 is the equivalence-based TTP method under $\al_{h2}=0.05$, and EQ2 is the equivalence-based TTP method under $\al_{h2}=0.10$}
  \end{center}
\end{figure}

\begin{figure}[H]
  \begin{center}
  \includegraphics[width=15cm]{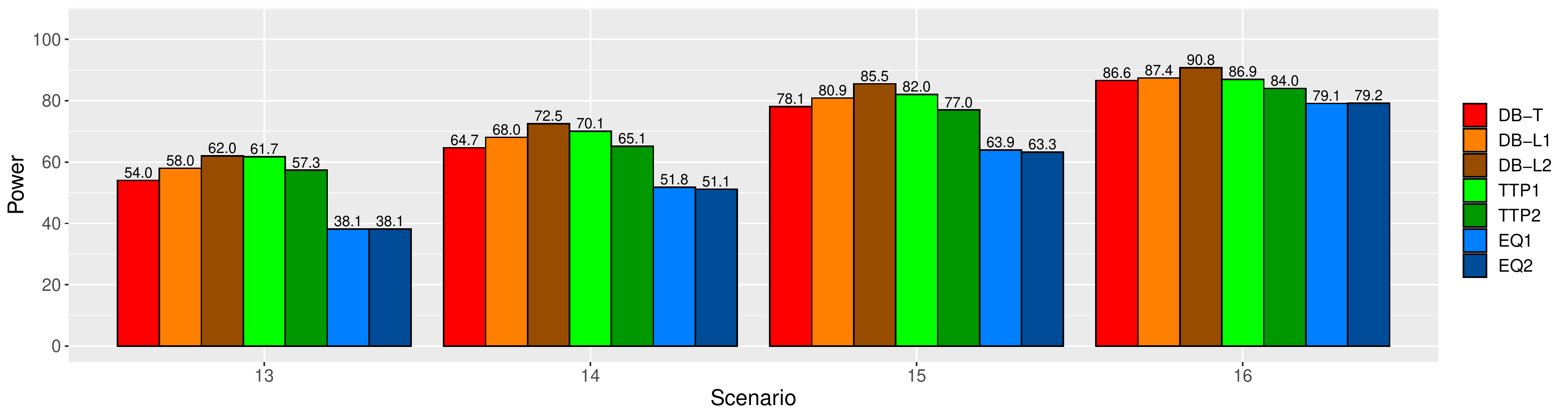}
  \caption{Power in Scenarios 13-16}
  \label{f:power3}
     \footnotesize{DB-T is the dynamic borrowing method based on the density function of the t-distribution, DB-L1 is the dynamic borrowing based on the logistic function with $\be_0=-7.379$ and $\be_1=4.472$, DB-L2 is the dynamic borrowing based on the logistic function with $\be_0=-7.374$ and $\be_1=3.747$, TTP1 is the original TTP method under $\al_{h1}=0.05$, TTP2 is the original TTP method under $\al_{h1}=0.10$, EQ1 is the equivalence-based TTP method under $\al_{h2}=0.05$, and EQ2 is the equivalence-based TTP method under $\al_{h2}=0.10$}
  \end{center}
\end{figure}

\begin{figure}[H]
  \begin{center}
  \includegraphics[width=15cm]{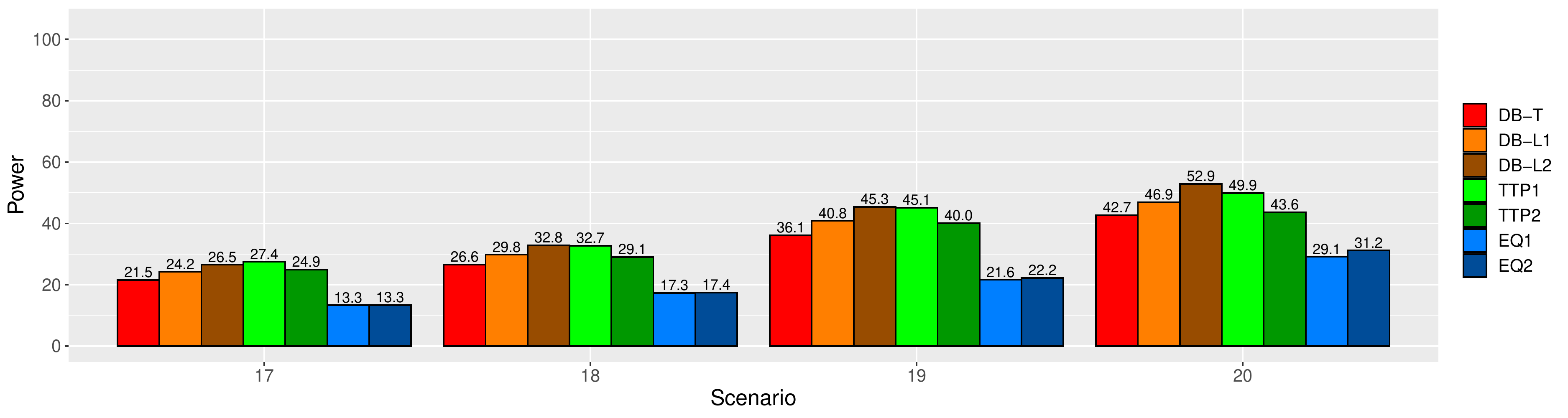}
  \caption{Power in Scenarios 17-20}
  \label{f:power4}
     \footnotesize{DB-T is the dynamic borrowing method based on the density function of the t-distribution, DB-L1 is the dynamic borrowing based on the logistic function with $\be_0=-7.379$ and $\be_1=4.472$, DB-L2 is the dynamic borrowing based on the logistic function with $\be_0=-7.374$ and $\be_1=3.747$, TTP1 is the original TTP method under $\al_{h1}=0.05$, TTP2 is the original TTP method under $\al_{h1}=0.10$, EQ1 is the equivalence-based TTP method under $\al_{h2}=0.05$, and EQ2 is the equivalence-based TTP method under $\al_{h2}=0.10$}
  \end{center}
\end{figure}

\begin{figure}[H]
  \begin{center}
  \includegraphics[width=15cm]{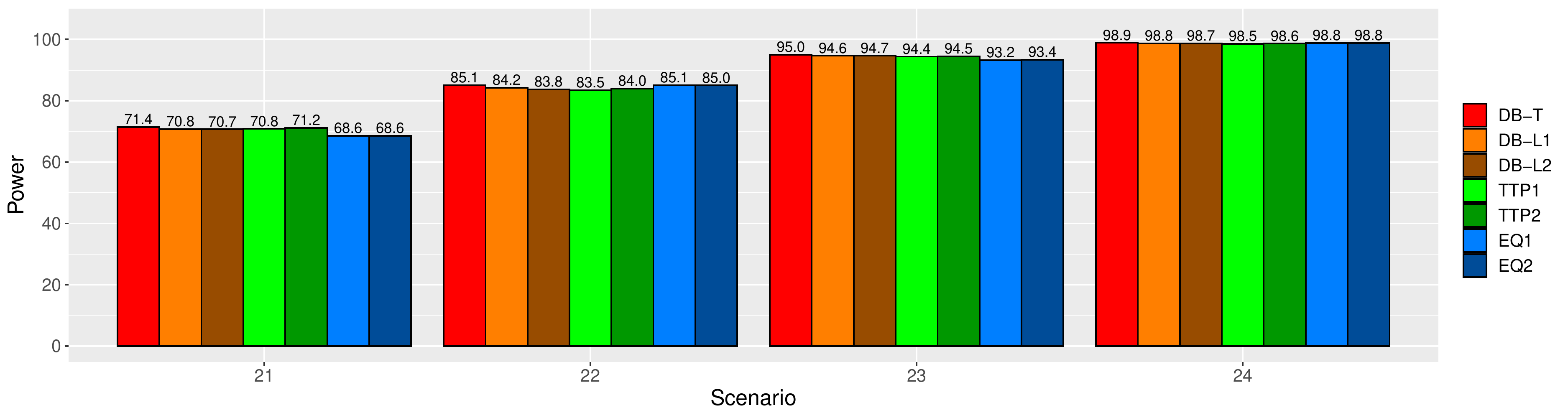}
  \caption{Power in Scenarios 21-24}
  \label{f:power5}
     \footnotesize{DB-T is the dynamic borrowing method based on the density function of the t-distribution, DB-L1 is the dynamic borrowing based on the logistic function with $\be_0=-7.379$ and $\be_1=4.472$, DB-L2 is the dynamic borrowing based on the logistic function with $\be_0=-7.374$ and $\be_1=3.747$, TTP1 is the original TTP method under $\al_{h1}=0.05$, TTP2 is the original TTP method under $\al_{h1}=0.10$, EQ1 is the equivalence-based TTP method under $\al_{h2}=0.05$, and EQ2 is the equivalence-based TTP method under $\al_{h2}=0.10$}
  \end{center}
\end{figure}

The power results for Scenarios 5-24 are shown in Figures \ref{f:power1}-\ref{f:power5}. For Scenarios 5-8 in Figure \ref{f:power1}, with the exception of scenario 5, DB-L2 had the highest power. EQ had lower detection power than the other methods. The difference decreases as the sample size increases.

For Scenarios 9-12 in Figure \ref{f:power2}, the powers of all methods were almost identical. The method with the highest detection power was EQ. DB-T was the most powerful of the proposed methods.

For Scenarios 13-16 in Figure \ref{f:power3}, DB-L2 had the highest power.

For Scenarios 17-20 in Figure \ref{f:power4}, with the exception of scenario 5, DB-L2 had the highest power.

For Scenarios 21-24 in Figure \ref{f:power5}, the results were almost identical, but DB-T had the highest detection power in all scenarios.

\section{Case studies}
\subsection{Real studies of major depressive disorder}
We reanalyzed real studies 059 (history) and 061 (current) in Keller et al.\cite{keller2006lack}. These studies were phase III clinical trials of patients with major depressive disorder treated using aprepitant or paroxetine and placebo. The primary efficacy endpoint was the mean change in scores from baseline for the first 17 items of the Hamilton Rating Scale for Depression at Week 8. The primary endpoint was not achieved. The second efficacy endpoint was the Hamilton Rating Scale for Anxiety (HAM-A) score. Although the HAM-A score did not shown a statistically significant difference, it should be noted that the sample may have been insufficient.

We analyzed the data in Li et al.~\cite{li2020revisit}. For the paroxetine 20 mg (current active comparator) group in study 061, the mean change in HAM-A score between baseline and Week 8 is -9.9, the standard deviation is 7.9, and the sample size is 137. For the placebo (current control) group in study 061, the mean change in HAM-A score between baseline and Week 8 is -8.7, the standard deviation is 7.3, and the sample size is 140. Efficacy was assessed at a one-sided significance level of 5\%. The difference between the paroxetine 20 mg and placebo groups (paroxetine - placebo) is not significant (-1.2; p = 0.0947); this may be attributable to an insufficient sample size. For the placebo (current control) group in study 059, the mean change between baseline and Week 8 in the HAM-A score is -8.1, the standard deviation is 8.3, and the sample size is 149.

We then apply our proposed dynamic borrowing method. For DB-T, the test statistic $T(A_t(-8.7,7.3,-8.1,8,3))$ is -1.81 and the critical value $z^\ast_{0.05}$ is -1.73 (p = 0.0408). For DB-L1, the test statistic $T(A_{l1}(-8.7,7.3,-8.1,8,3))$ is -1.85 and the critical value $z^\ast_{0.05}$ is -1.72 (p = 0.0378). For DB-L2, the test statistic $T(A_{l2}(-8.7,7.3,-8.1,8,3))$ is -1.85 and the critical value $z^\ast_{0.05}$ is -1.70 ( p = 0.0364). All p-values were under $0.05$; therefore, the application of our proposed methods resulted in statistically significant differences. A summary of the reanalysis is shown in Table \ref{t:a_sum}.

\begin{table}[H]
  \begin{center}
\caption{Summary of actual analysis\label{t:a_sum}}
\begin{tabular}{|c|c|c|c|}\hline
& \multicolumn{2}{c|}{Study 061} & Study 059 \\\hline
N & 137 & 140 & 149 \\
mean & -9.9 & -8.7 & -8.1 \\
SD & 7.9 & 7.3 & 8.3 \\\hline\hline
 & DB-T & DB-L1 & DB-L2 \\\hline
Borrowing level & 0.81 & 0.99 & 0.99\\
Test statistic & -1.81 & -1.85 & -1.82\\
$z^\ast_{0.05}$ & -1.73 & -1.72 & -1.70 \\
p-value$^\ast$ & 0.0408 & 0.0378 & 0.0364 \\\hline
\end{tabular}
\\
\footnotesize{$z^\ast_{0.05}$ is the lower 5\% percentile of the 10,000 bootstrap sets. The p-value$^\ast$ is the proportion of the 10,000 test statistics computed from the 10,000 bootstrap sets that are smaller than the test statistic calculated from the original data.}
  \end{center}
\end{table}

\section{Discussion}
We proposed a dynamic borrowing method for historical information using a frequentist approach. Our method borrows 0\% to 100\% of historical data on the basis of the density function of the t-distribution or the logistic function. The pooled mean and variance of current and historical control data are calculated according to the borrowing amount. The test statistic of our proposed method is very easy to obtain (as a Z-test statistic) using the pooled mean and variance. However, as a limitation, the asymptotic distribution of the test statistic cannot be expressed in closed form. Hence, a parametric bootstrap method is needed. The simulation results confirm excellent results of the bootstrap method compared with the conventional method.

In the simulation studies, our proposed methods controlled the type 1 error rate at a prespecified nominal significance level. Moreover, our methods had higher power than previous methods. The method based on the density function of the t-distribution had more power than the other methods in cases where the historical control group was better than the current control group (scenarios 9-12 and 21-24, where we did not want to excessively borrow information). The power of the method based on the logistic function showed a similar trend to that of the original TTP method in all scenarios. However, in most scenarios, the proposed method was more powerful than the original TTP method and more robust to type 1 error. For the original and equivalence-based TTP methods, type 1 error tended to inflate slightly with decreasing sample size. The original and equivalence-based TTP methods used an approximation to the Z distribution to control type 1 error. In practice, the statistics used in the original and equivalence-based TTP methods follow a t-distribution. Thus, we considered the type 1 error rates to be slightly inflated when the sample size was small. However, as the bias of bootstrap method depends only on the number of bootstrap samples, we consider that the type 1 error rates for our proposed method were well controlled by including a sufficiently large number of bootstrap samples even when the sample size was small. We prepared the R Shiny application to execute a simulation study with our proposed methods. The simulation application is available online \href{https://masa-koji.shinyapps.io/D-BT/}{Simulation of the dynamic borrowing method}.

In the reanalysis of actual studies, the information borrowing level of our proposed method based on the t-distribution was 0.81. Because of a slight difference in the means, about 20\% of the historical control information was not borrowed. The information borrowing level of our proposed method based on the logistic function was 0.99, i.e., almost all historical data were used. Information borrowing showed a statistically significant difference among all methods. We only used summary statistics in the analysis of actual data. Our proposed method allowed us to analyze cases both where individual data were available and were unavailable.

We confirmed that our proposed methods effectively controlled type 1 error compared with the conventional original and equivalence-based TTP methods, even when the sample size was small. In addition, our proposed methods had higher statistical power than the conventional original and equivalence-based TTP methods. Therefore, there were no problems with our proposed methods. In particular, we recommend the method based on the logistic function, which had consistently high statistical power. In the future, we will develop an optimal parameter estimation method for the logistic function.

\bigskip
\noindent
{\bf Acknowledgements.}
MK would like to thank Professor Hisashi Noma for his encouragement and helpful suggestions.

\bibliography{main.bib} 
\bibliographystyle{unsrt} 

\end{document}